\documentclass[10pt]{iopart}
\usepackage{iopams} 

\usepackage{hyperref}
\usepackage{graphicx}
\usepackage{dcolumn}
\usepackage{bm}
\usepackage{comment}
\usepackage{float}
\usepackage{caption}
\usepackage{subfig}

\begin{document}
\title{Pedestal and edge turbulence characteristics from an XGC1 gyrokinetic simulation}
\author[1]{R.M. Churchill$^1$, C.S. Chang$^1$, S. Ku$^1$}
\address{$^1$ Princeton Plasma Physics Laboratory, 100 Stellarator Road, Princeton, NJ 08540, USA}
\ead{rchurchi@pppl.gov}

\begin{abstract}
Understanding the multi-scale neoclassical and turbulence physics in the edge region (pedestal + scrape-off layer) is required in order to reliably predict performance in future fusion devices. We explore turbulent characteristics in the edge region from a multiscale neoclassical and turbulent XGC1 gyrokinetic simulation in a DIII-D like tokamak geometry, here excluding neutrals and collisions. For an H-mode type plasma with steep pedestal, it is found that the electron density fluctuations increase towards the separatrix, and stay high well into the SOL, reaching a maximum value of $\delta n_e / \bar{n}_e \sim 0.18$. Blobs are observed, born around the magnetic separatrix surface and propagate radially outward with velocities generally less than 1 km/s. Strong poloidal motion of the blobs is also present, near 20 km/s, consistent with $E \times B$ rotation. The electron density fluctuations show a negative skewness in the closed field line pedestal regions, consistent with the presence of "holes", followed by a transition to strong positive skewness across the separatrix and into the SOL. These simulations indicate that not only neoclassical phenomena, but also turbulence, including the blob-generation mechanism, can remain important in the steep H-mode pedestal and SOL. Qualitative comparisons will be made to experimental observations.
\end{abstract}

\maketitle
\ioptwocol

\section{Introduction}
The edge region of diverted tokamak plasmas is generally recognized as critical to the performance and success of a fusing plasma. First, it is desirable to maximize the H-mode\cite{Wagner1984} pressure pedestal at the edge of the closed field line regions, since this sets the amount of fusion power in a given size machine\cite{Doyle2007} due to core gradients being limited by turbulence. Second, it is desirable to minimize plasma-material interaction in the open-field line region immediately outside of the confined plasma (the scrape-off layer), to prevent wall erosion and avoid impurities and neutrals from limiting plasma performance.

Since these two edge regions (pedestal and SOL) are considered to be important for different reasons, they are often treated as separable in both experiment and simulation studies. However, it's also long been recognized that certain physics couple these two regions, such as X-point ion orbit loss\cite{Chang2002} and nonlocal turbulence dynamics, including blob formation and propagation. To gain a holistic understanding of fusion edge plasmas using simulation, it is advisable to include in simulation codes as much of the relevant physics as possible and needed.

The goal of the gyrokinetic XGC1 code is exactly this, to include the physics necessary to correctly model the edge, including the pedestal region and scrape-off layer without a scale-separation assumption. XGC1 is a total-$f$, gyrokinetic particle-in-cell (PIC) code, which includes relevant edge physics such as turbulence, neoclassical physics, experimental magnetic equilibrium including X-points, realistic divertor geometry, Monte Carlo neutrals, and self-consistent calculation of the background electric field profile.

While simultaneously including all of this physics gives the benefit of more realistic simulations, it also brings the challenge of extracting or isolating the dominant processes. In this sense, XGC1 is similar to experiments, where a series of diagnostic outputs need to be processed and analyzed to extract the salient physics.

The purpose of this paper is to study the turbulence characteristics in the edge from a sample XGC1 simulation, using techniques commonly applied to plasma experiments. This study will highlight unique features of edge transport and turbulence as seen from this example XGC1 simulation in a DIII-D-like H-mode plasma, which is a difficult region to cover diagnostically in experiments.  Generic comparisons to published experimental results will be pointed out.

Section \ref{sec:xgc1} describes the simulation inputs and parameters of the XGC1 simulation. Section \ref{sec:turb} calculates basic turbulence properties from XGC1 density fluctuations, including the skewness and kurtosis relationship of the fluctuations and the conditional spectrum of the density fluctuations across the pedestal region. Section \ref{sec:blobs} applies a blob tracking technique to characterize propagation of blobs and their internal potential structure. Section \ref{sec:discussion} discusses future edge turbulence topics to explore with the XGC1 code. Finally, Section \ref{sec:conclusion} concludes with a summary of results presented.

\section{XGC1 simulation} \label{sec:xgc1}
\begin{figure}
\centering
\includegraphics[width=0.45\textwidth]{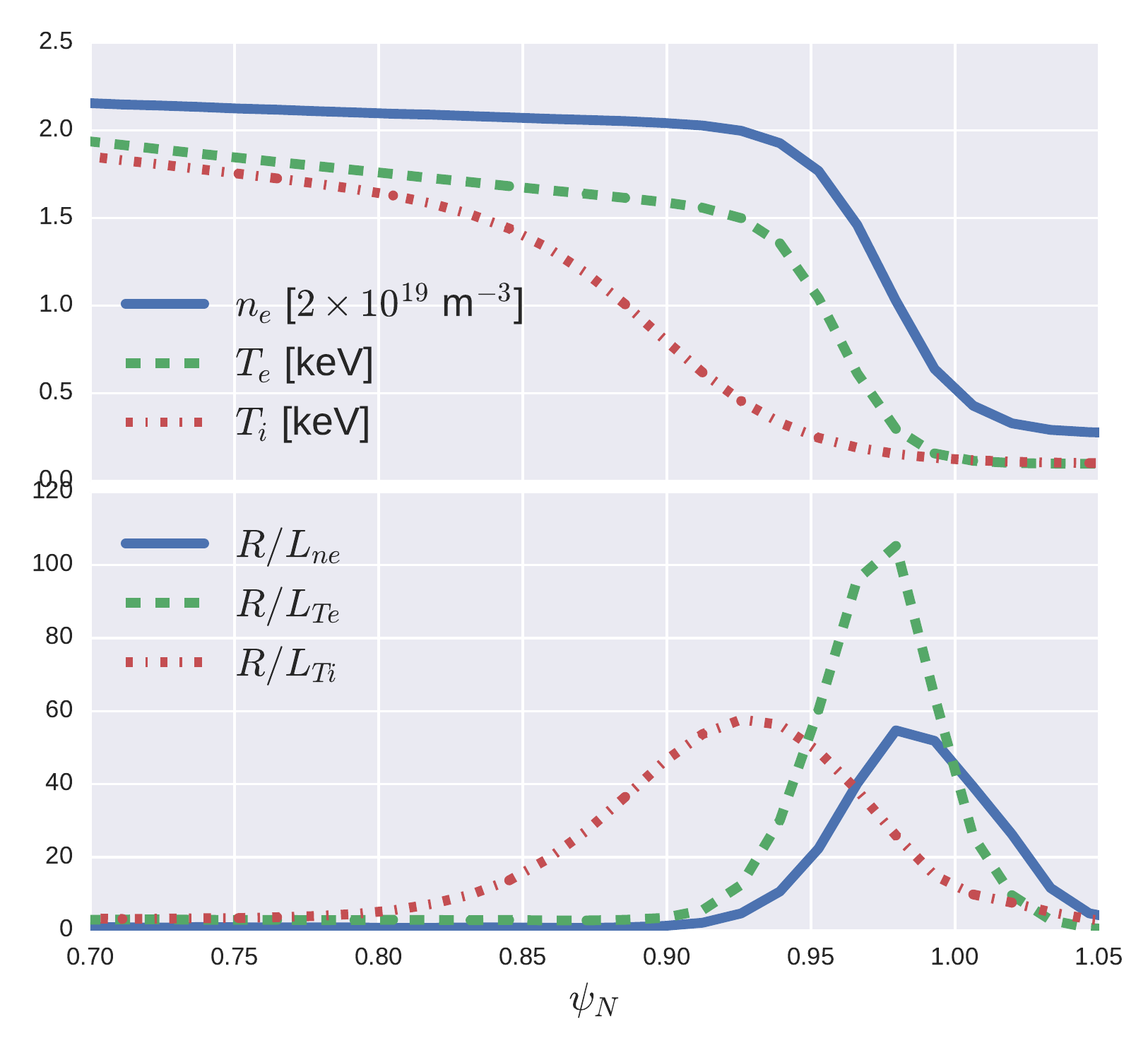}
\caption{Model input profiles of electron density, temperature, and ion temperature. Also plotted are the normalized gradient scale lengths, $R/L$, with $L_y = y \cdot |dy/dr|^{-1}$}
\label{fig:inputs}
\end{figure}

The inner-workings of the XGC1 code won't be detailed here, the interested reader can find more details in several references\cite{Ku2017,Ku2016,Ku2009}. In the present simulation, we use an electrostatic version of XGC1. This version also utilizes the hybrid-Lagrangian total-$\delta f$ scheme\cite{Ku2016}, which utilizes a phase space grid for the coarse-grained, slow time-varying particle distribution function, $f_0 = f - \delta f$, combined with the usual marker particles of a PIC simulation to hold a $\delta f$ piece of the distribution function. Both $f_0$ and $\delta f$ are evolved together according to the gyrokinetic equations, and no assumption of the size of $\delta f$ relative to $f_0$ is made, which is crucial in the edge where $\delta f$ can be the same order as $f_0$. Since $f_0$ is numerically solved under wall loss and charge-exchange and ionization with neutral particles, it can be highly non-Maxwellian. Gyrokinetic ions and drift-kinetic electrons are used. The present study uses a realistic DIII-D-like magnetic equilibrium, including lower single-null X-point, with $q_{95} \approx 3.5$ and $B_0\approx 2.1$ T. Collisions and neutrals were turned off in this particular report, in order to remove the Coloumb collision and neutral particle effects for simplicity, though these are typically used in divertor heat-flux studies in Refs. \cite{Chang2017} and \cite{Chang2017a}). Future reports will study these effects on blobby turbulence one by one. Model H-mode input profiles of the electron density ($n_e$), electron temperature ($T_e$), and ion temperature ($T_i$) were used, see Figure \ref{fig:inputs}. Note the relative shift in the gradient of $n_e$, $T_e$, and $T_i$.

XGC1 utilizes an unstructured, field-aligned mesh for field solves and cylindrical ($R,\phi,Z$) coordinates for particle time advances. The mesh covers the entire plasma up to a pre-defined flux-surface, typically the last flux surface in the far-SOL not intercepted by a limiter, but still including the divertor plates in the simulation domain. The electrostatic potential is approximated to zero for outside of this far-SOL flux surface. Data output such as the moments of the distribution function and the electrostatic potential are output at the vertices of the unstructured mesh, at a number of poloidal planes (in this simulation 16). The grid spacing in this simulation has a $\Delta R$ at the low-field side midplane of 3mm, and a poloidal arc length averaging about 7mm on the separatrix flux-surface.

The time-step used in the simulation was 0.15 $\mu$s.

\section{Turbulence properties}\label{sec:turb}
\begin{figure}
\centering
\includegraphics[width=0.45\textwidth]{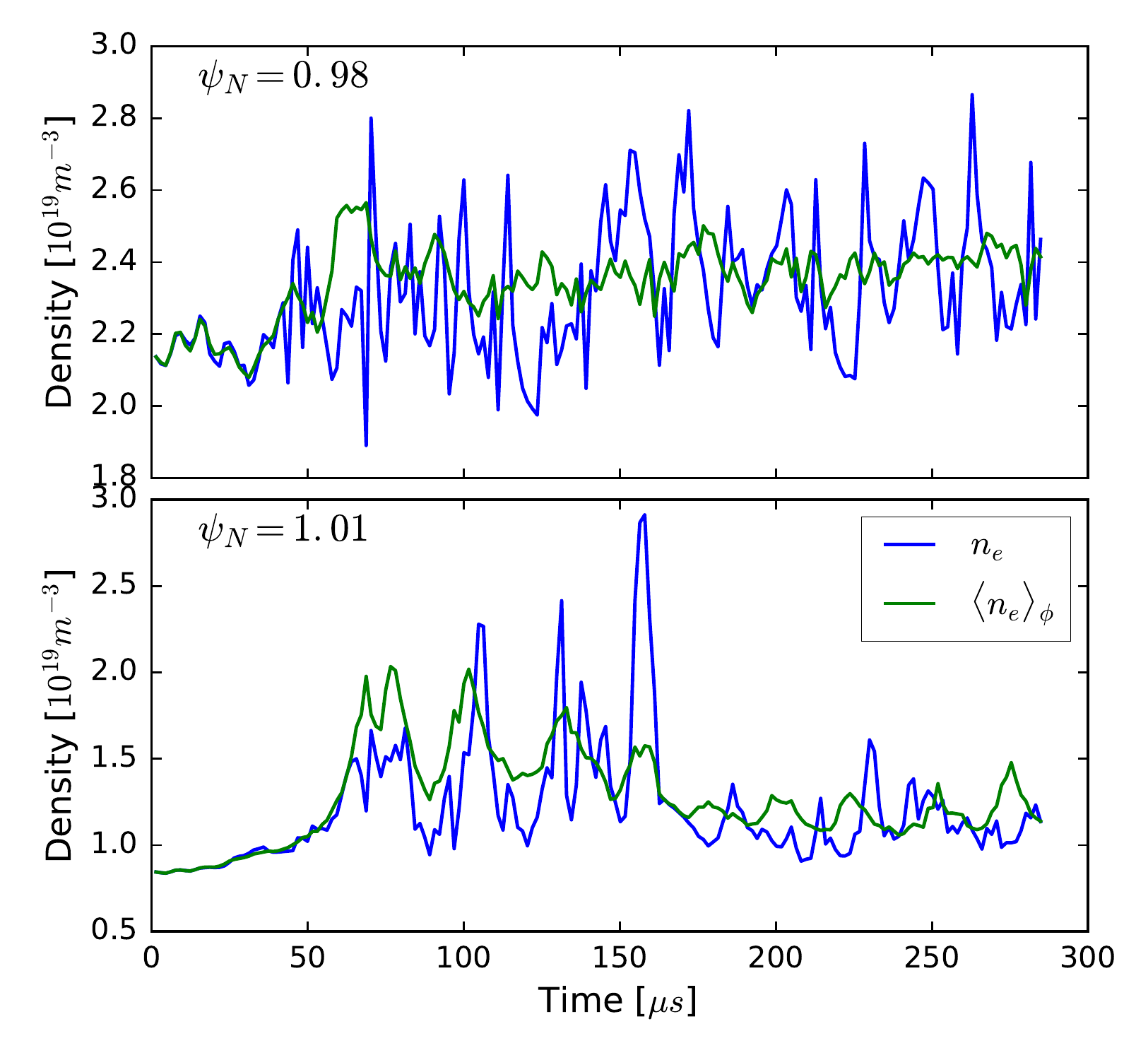}
\caption{Fluctuating density traces in the pedestal and SOL}
\label{fig:ne_time}
\end{figure}

For the sake of discussion in this paper, we define the ''near-SOL" to be $1<\psi_N<1.01$, where $\psi_N$ is the normalized poloidal flux, $\psi_N = (\psi - \psi_0)/(\psi_x - \psi_0)$, with $\psi_0$ the poloidal flux at the magnetic axis, and $\psi_x$ the poloidal flux at the separatrix. This near- and far-SOL definition is in relation with the divertor heat-load width mapped to the outboard midplane that is $< 4$ mm or $\Delta \psi_N =0.01$ for most of the DIII-D discharges \cite{Eich2013}. Time traces of the electron density from this XGC1 simulation at two spatial points at the LFS midplane are shown in Figure \ref{fig:ne_time}, one in the middle of the pedestal ($\psi_N=0.98$) and the other at the boundary (in our definition) of the near- and far-SOL($\psi_N=1.01$). Large amplitude, intermittent burst are observed in both regions, but the SOL time trace has even larger amplitude bursts above it's baseline density. The toroidally averaged density is also plotted, $\langle n_e \rangle_\phi = \int d\phi \, n_e / \int d\phi$, showing that locally the toroidally averaged electron density decreases from the initial value due to loss to divertor plates.  Outward particle transport from  inside the separatrix surface slows down the density decrease at $\psi_N=1.01$. The initial transient phase before turbulence saturates (up to around 100 $\mu$s) is dominated by geodesic acoustic modes (GAM's), a common feature of total-$f$ simulations, due to the initial marker particle distribution relaxing from a Maxwellian at the time scale of a few ion toroidal transit times. After nonlinear saturation of the turbulence ($t \ge 100 \mu$s), a residual GAM oscillation is present, with a characteristic frequency near 40 kHz (the GAM-like oscillations in the SOL may be due to nonlocal effects from the closed field line region). Experimentally, GAM's have not been observed in H-mode plasmas\cite{Conway2005}. Such a residual GAM oscillation is not surprising in the present short time, collisionless simulation: the strong collision damping mechanism is absent and only the weak Landau-resonance damping exists. In order to isolate turbulent characteristics of interest, in the rest of this paper the nonlinear, saturated turbulent state ($t \ge 100 \mu$s) will be analyzed, with the toroidally averaged electron density ($\langle n_e \rangle_\phi$) removed from the total density ($\delta n_e = n_e - \langle n_e \rangle_\phi$).

We begin then by investigating the basic turbulence characteristics from this XGC1 simulation, including the normalized fluctuation level, autocorrelation time, and correlation lengths (radial and poloidal) of the electron density fluctuations. These characteristics are binned over a poloidal wedge region located at the low-field side, spanning all toroidal angles and a poloidal angle extent of roughly $\left| \theta  \right| < 40^\circ$.

\subsection{Normalized fluctuation level}
\begin{figure*}[ht]
\centering
\subfloat{\label{fig:dneOverne0} \includegraphics[width=0.45\textwidth]{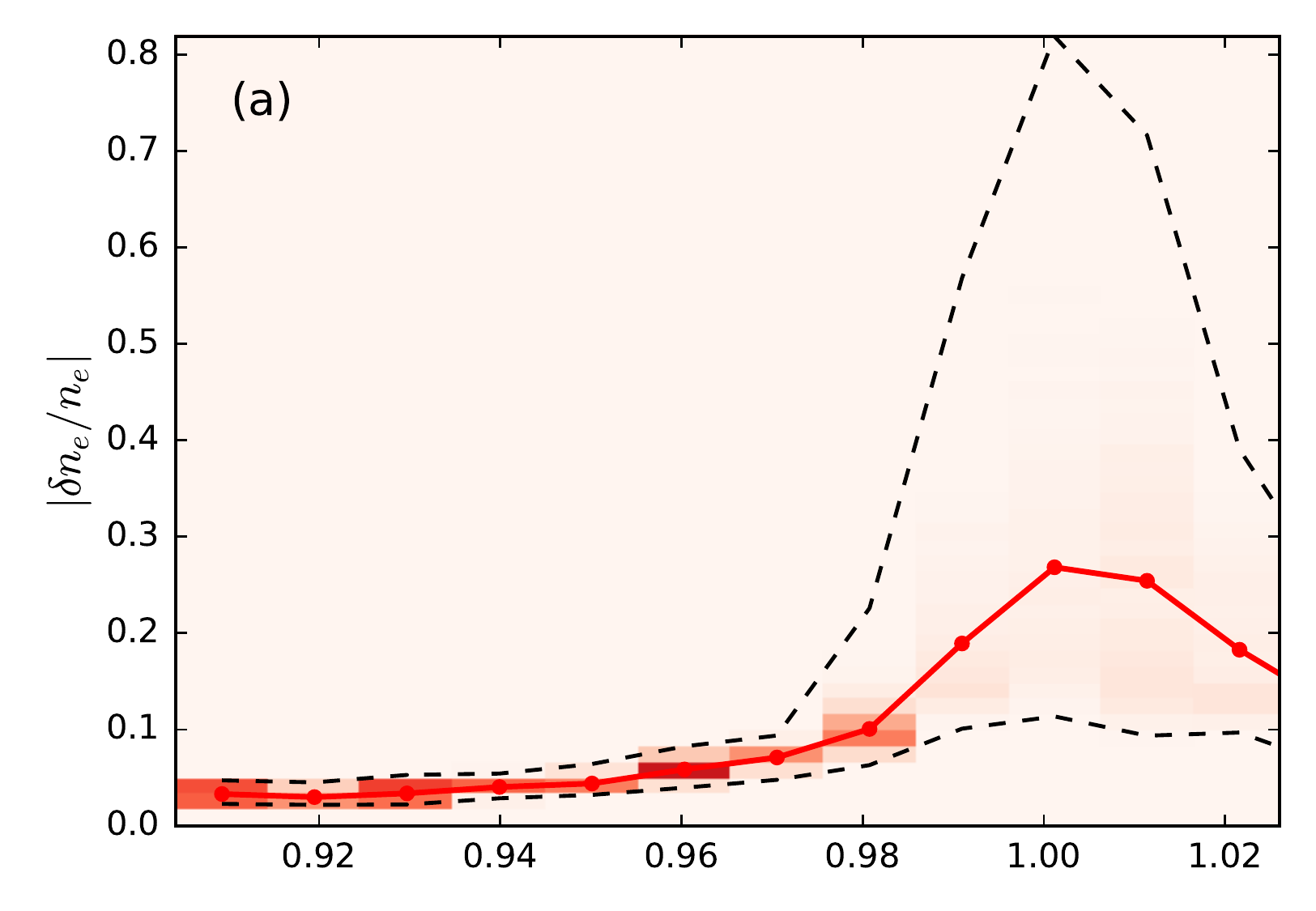}        }
\enskip
\subfloat{\label{fig:tauac} \includegraphics[width=0.45\textwidth]{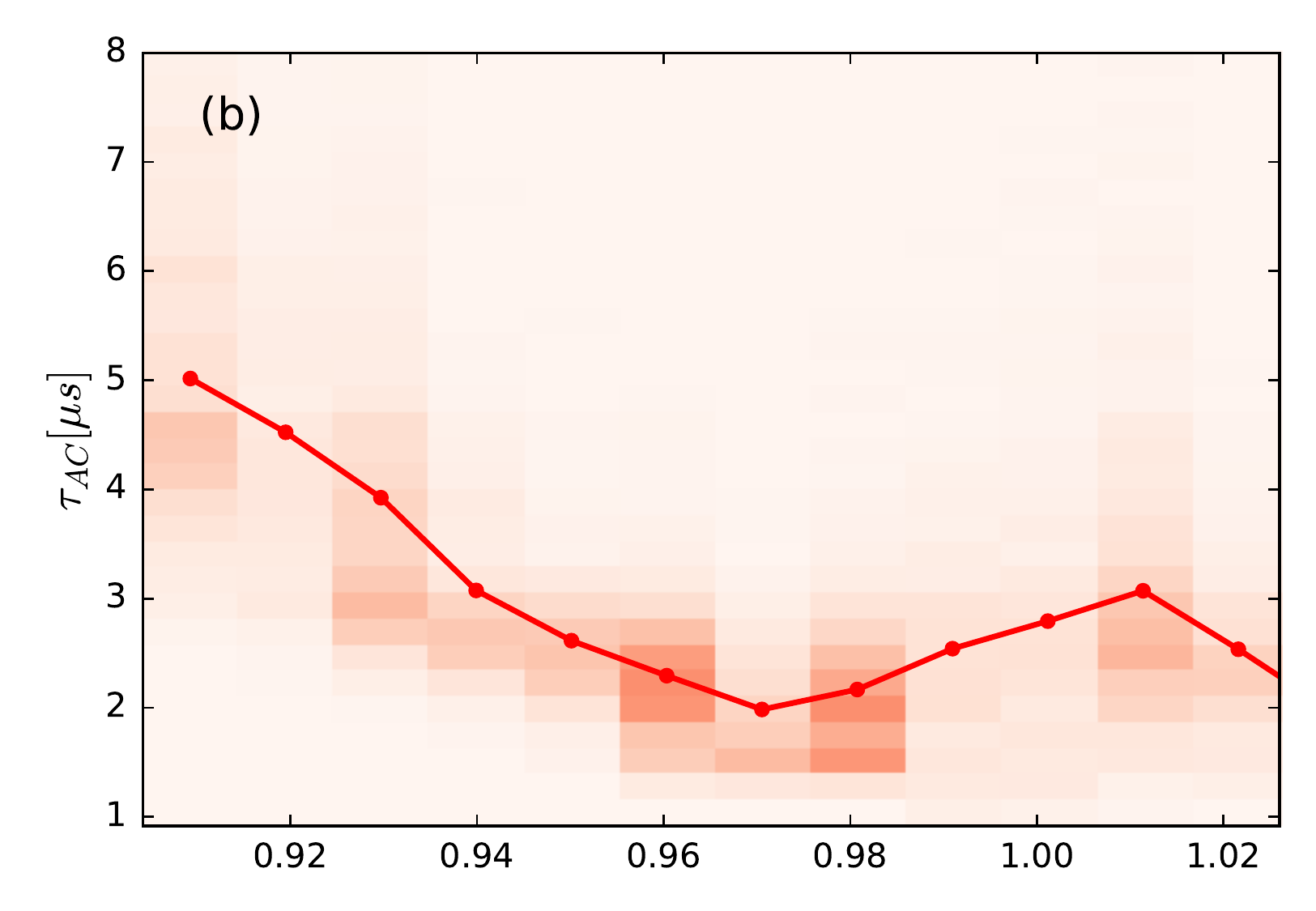}        }
\vspace{0.1in}
\subfloat{\label{fig:Lrad} \includegraphics[width=0.45\textwidth]{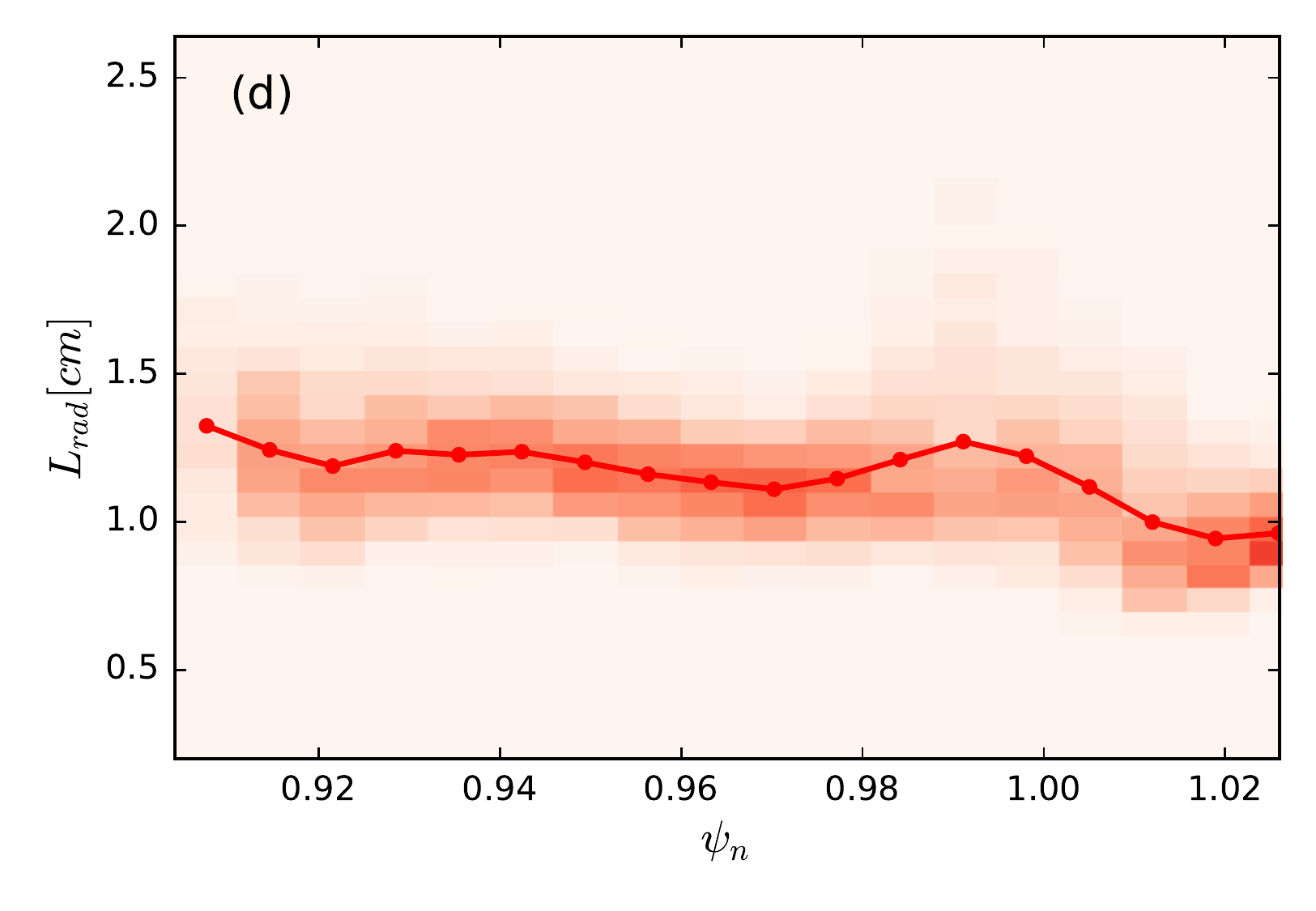}}
\enskip
\subfloat{\label{fig:Lpol} \includegraphics[width=0.45\textwidth]{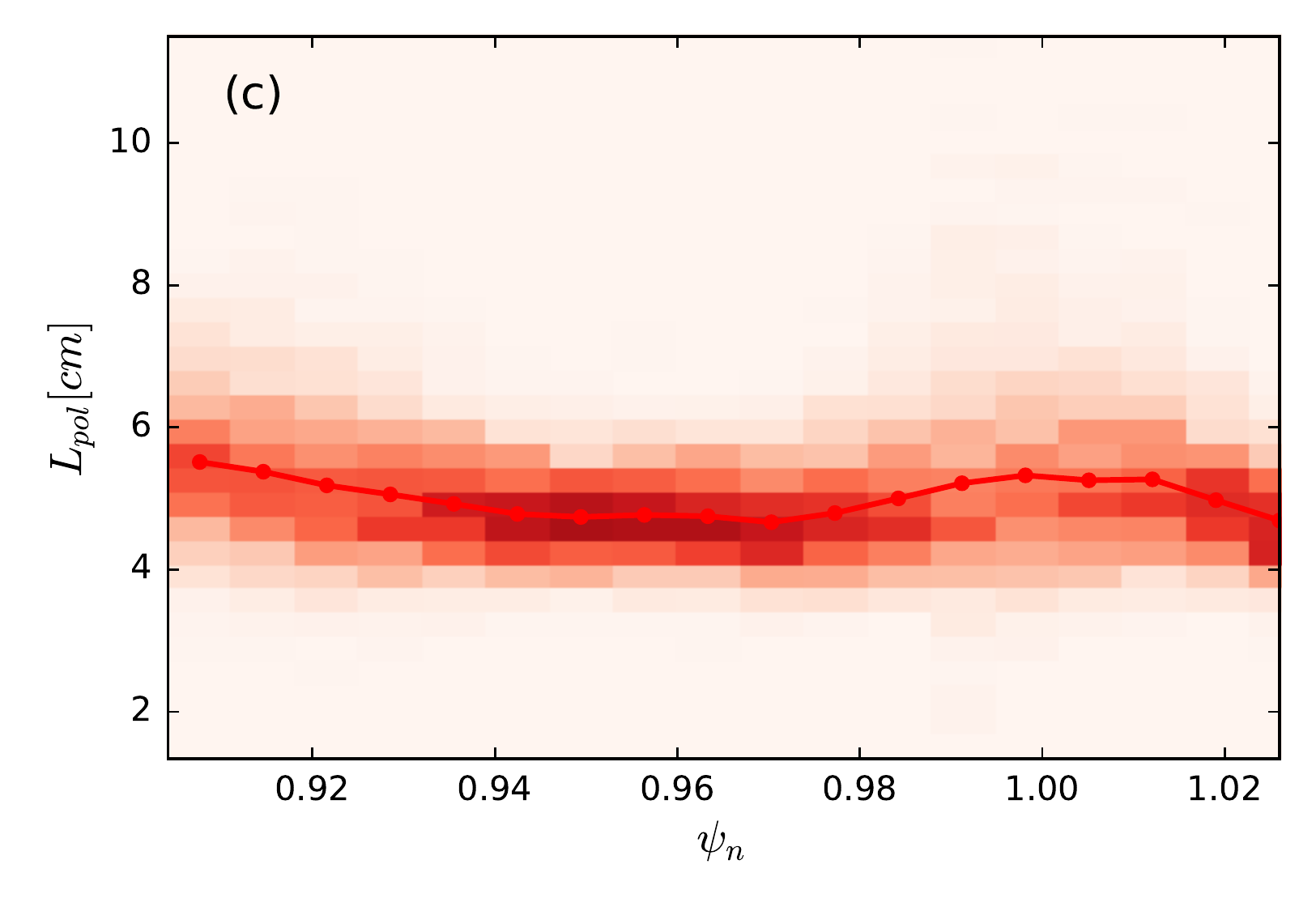}        }
\caption{Turbulence characteristics of the fluctuating electron density, binned into a histogram (color contours),then averaged (red line with circle symbols) over space centered at LFS midplane, extending $\left| \theta \right| < 40$ (a) normalized fluctuating electron density $\delta n_e / n_e$, (b) autocorrelation time, (c) radial correlation length, and (d) poloidal correlation length. Black dashed lines in (a) represent maximum and minimum over the entire poloidal wedge region. }
\end{figure*}

The normalized fluctuation level $\left| \delta n_e / \langle n_e \rangle_\phi \right|$ is shown in Figure \ref{fig:dneOverne0}, with the average, minimum, and maximum plotted as described in the caption. This normalized fluctuation level is calculated by taking the standard deviation over time of the fluctuating electron density, $\delta n_e = n_e - \langle n_e \rangle_\phi $, and normalizing to the mean over time of the toroidally averaged density, $\langle \langle n_e \rangle_\phi \rangle_t$. Figure \ref{fig:dneOverne0} is a 2d histogram of this quantity over the poloidal wedge region. The value of $\left | \delta n_e/ \langle n_e \rangle_\phi \right|$ near the top of the pedestal is around 4\%, with little spatial variation over the wedge region. A gradual increase is seen through most of the pedestal region, until $\psi_N \sim 0.98$, where there is a sharp increase. This location of the beginning of a sharp increase in $\left| \delta n_e / n_{e0} \right|$ also coincides with the peak in the inverse gradient scale length of the density, $L_{n}^{-1} = | \nabla n/n |$. The normalized density fluctuations continue rising through the transition across the separatrix, reaching a peak of ${\sim}30\%$ and then decrease gradually into the far scrape-off layer. Already we begin to see the signs of intermittency in the region spanning the foot of the pedestal to the middle of the scrape-off layer ($0.98 < \psi_N < 1.025$), as the mean value of $|\delta n_e/n_{e0}|$ lies much closer to the minimum value, implying a high skewness. We will explore this more in depth in the sections on skewness and blob transport.

\subsection{Autocorrelation Time}
The following correlation analysis makes use of the correlation function\cite{Simon2014}:

\begin{equation}
\centering
C_{XY}(\tau) = \frac{1}{\sigma_X \sigma_Y} \int dt X(t+\tau) Y(t)
\end{equation}

\begin{figure}
\centering
\includegraphics[width=0.45\textwidth]{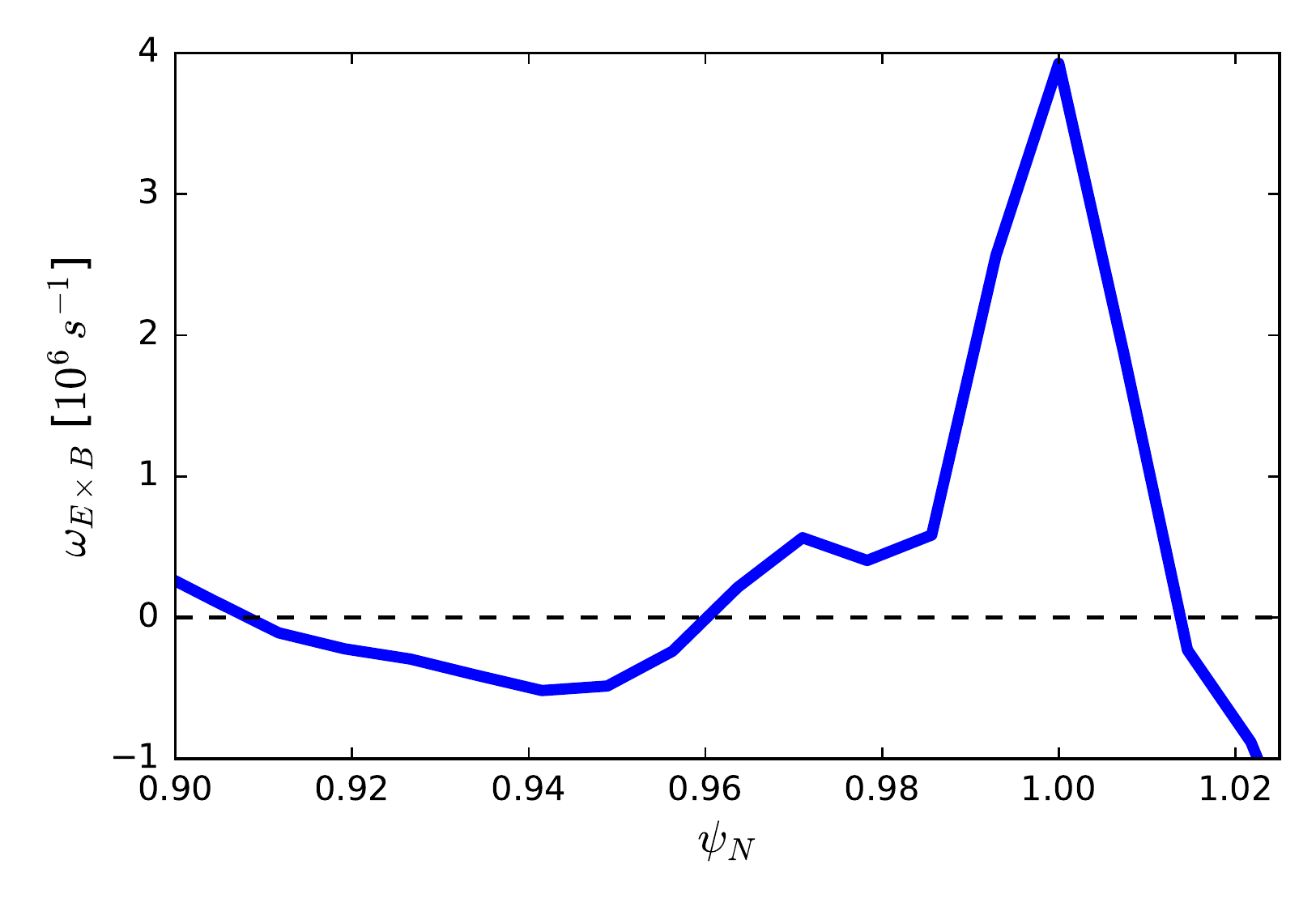}
\caption{Time averaged $E \times B$ shearing rate at the LFS midplane. }
\label{fig:wexb}
\end{figure}

The autocorrelation time gives an indication of the eddy turnover time. $\tau_{ac}$ is calculated by finding the $1/e$ point of the autocorrelation function, $C_{XX}(\tau)$, with $X=\delta n_e$\cite{Simon2014}. Figure \ref{fig:tauac} shows the autocorrelation time binned over the spatial points in the LFS wedge region. The reduction in $\tau_{ac} $from $0.92<\psi_N<0.96$ corresponds to a region of negative $E \times B$ shearing rate\cite{Hahm1995}, $\omega_{E \times B} = r/q \, \partial_r \left(q E_r / rB \right)$, seen in Figure \ref{fig:wexb}, suggesting that turbulent eddies are being sheared apart by $E \times B$ flows. Interestingly, the autocorrelation time increase towards the separatrix, even though the $E \times B$ shearing rate grows very large there. At first impression, this would seem to support several findings suggesting that the inner leg of the radial electric field is the most effective for turbulent transport reduction\cite{Viezzer2014}. It should be noted however, the autocorrelation time itself can be dominated by $E \times B$ and diamagnetic motions\cite{Endler1995}, and therefore may not be the best indication of the effectiveness of $E \times B$ shear decorrelating eddies, thereby reducing transport. Still, the radial profile of the relative density fluctuation level and the auto-correlation in Figs. \ref{fig:dneOverne0} and \ref{fig:tauac} implies that the turbulence property could be different between $\psi_N < 0.98$ and $> 0.98$.

\subsection{Correlation Lengths}
Correlation lengths give us an indication of the turbulent eddy size, and are quantities often used in diagnosing experiments, especially when there sufficient time resolution in a diagnostic but less spatial resolution. We will compare these calculations to the image processing results in Section \ref{sec:blobs}. Radial and poloidal correlation lengths ($L_{r}$ and $L_\theta$) of the electron density fluctuations are calculated over the LFS wedge region, and binned over all toroidal planes in Figure \ref{fig:Lrad} and Figure \ref{fig:Lpol}. For reference, at the LFS midplane a distance of 1 cm is approximately $\Delta \psi_N \sim 0.025$. Correlation lengths are calculated using the formula $L_x = 1.66 \Delta x / \sqrt{-\ln C_{12}}$, where $C_{12}$ is the correlation coefficient of the time series of two neighboring spatial points, and $\Delta_x$ is the radial or poloidal distance between the points\cite{Zweben2015}. For the radial and poloidal correlation lengths, points which are dominantly separated in the radial or poloidal direction are used respectively. The average poloidal correlation length from this simulation over most of the edge region is ${\sim}$5 cm, a size similar to $L_\theta$ measurements by BES is the core\cite{Mckee2005}. The average radial correlation length from this simulation is around 1.25 cm, a factor of ${\sim}$4x smaller than the poloidal correlation length. Typical experimental scalings of poloidal to radial correlation lengths are $L_\theta \sim (1-2) L_r$, although factors of ${\sim}$4 were observed on TEXTOR\cite{Zweben2007}. Measurements on DIII-D with the correlation reflectometer yielded values of edge $L_r$ in the range of 0.5 - 2 cm, mainly for L-mode but also for H-mode, where $L_r$ dropped from its L-mode values\cite{Rhodes2002}. Poloidal correlation lengths were measured inwards of the pedestal with beam emission spectroscopy (BES), with values in the range of 4-6 cm, in-line with the values in this simulation\cite{Mckee2005} (albeit not measured at the same plasmas). It is noteworthy in this simulation that across a region of varying gradients in background density, temperature, and electric fields, and across the transition to open field lines, that the radial correlation length of the turbulence remains fairly constant. We note again here that the present observations are from a collisionless plasma, while the experimental observations are from highly collisional plasmas.  How collisional dissipation affects the blobby turbulence properties is the subject of a subsequent paper.



\subsection{Skewness and kurtosis} \label{sec:skew}
\begin{figure}
\centering
\subfloat[]{\includegraphics[width=0.45\textwidth]{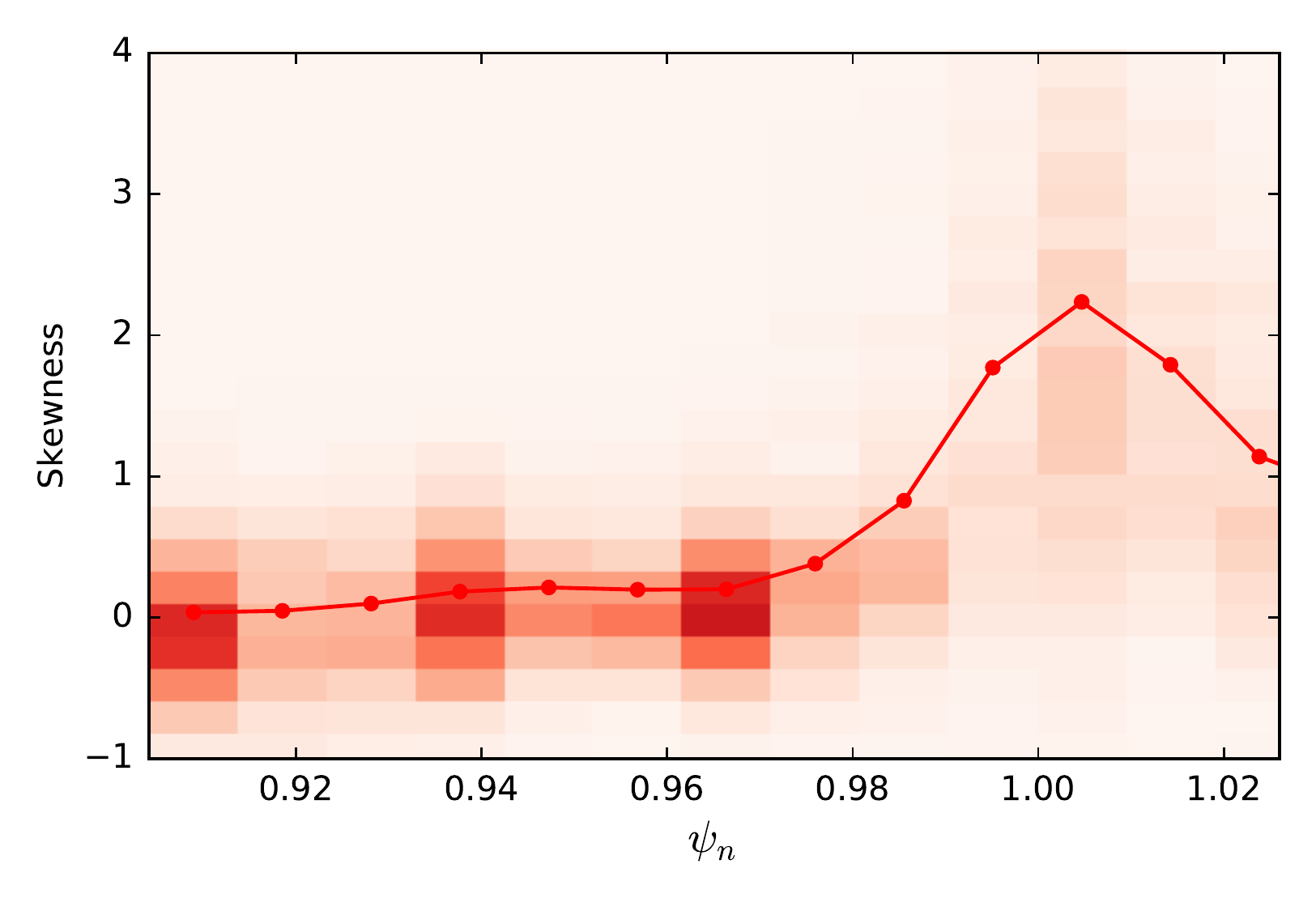} \label{fig:skew}}
\vskip\baselineskip
\subfloat[]{\includegraphics[width=0.45\textwidth]{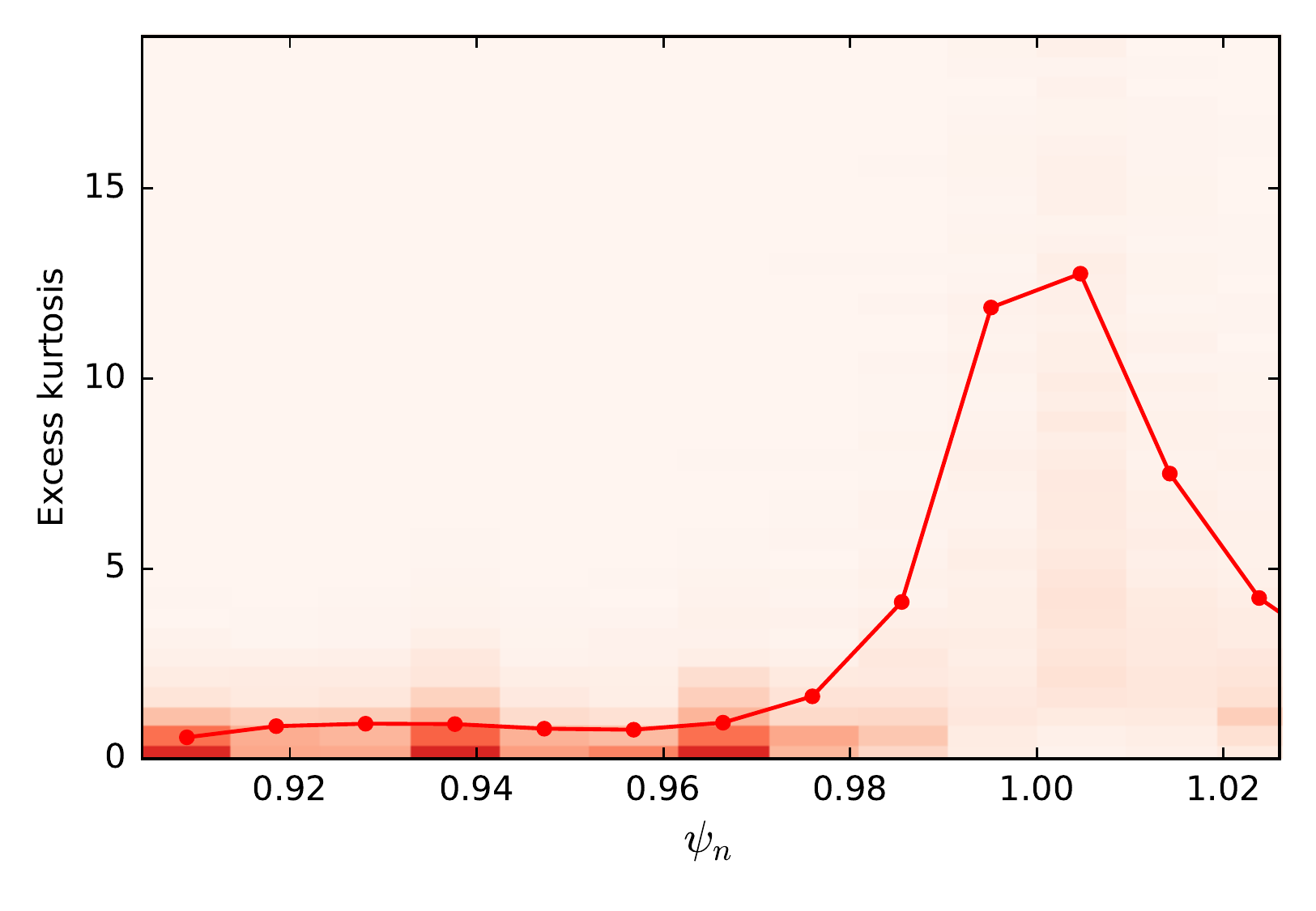} \label{fig:kurt}}
\caption{Skewness (\ref{fig:skew}) and kurtosis (\ref{fig:kurt}) of the density fluctuations in a region centered around the LFS midplane. }
\label{fig:skewkurt}
\end{figure}

It is common for turbulent fluctuations in the edge of magnetic fusion devices to have a decidedly non-Gaussian distribution\cite{Antar2003}. Calculating the third and fourth order moments of their distribution (skewness and kurtosis) helps identify the underlying distribution and identify processes which drive such features\cite{Labit2007,Krommes2008} (although it has been argued that in fact the universality of non-Gaussian features in edge simulations without full magnetic geometry, etc. would suggest such features are not a ''sensitive probe of the underlying physical mechansims''\cite{Krasheninnikov2008a}) . The skewness versus radial position in the plasma indicates the dominance of positive or negative departures from the mean (commonly referred to as ''blobs" and ''holes" in the edge region\cite{DIppolito2004}). The kurtosis gives a measure of the level of intermittency at that location. Figure \ref{fig:skewkurt} has plots of skewness and excess kurtosis of $\delta n_e$ versus $\psi_N$, again binned over the LFS wedge region described in the previous section. Similar to the radial variation of $\delta n_e/ \langle n_e \rangle_\phi$, the skewness and kurtosis increase significantly near the bottom of the $T_e$ pedestal, which also coincides with the steepest gradient in $n_e$. They both stay high into the near-SOL, decreasing into the far-SOL. The skewness near the top of the pedestal ($\psi_N < 0.92$) is slightly negative at many points, an observation similar to BES measurements\cite{Boedo2003} which implies the existence of ''holes" in this region. From the plot, we can see that throughout the edge region, the average excess kurtosis is above zero, meaning that throughout the edge region the fluctuations have non-Gaussian characteristics.

\begin{figure}
\centering
\includegraphics[width=0.45\textwidth]{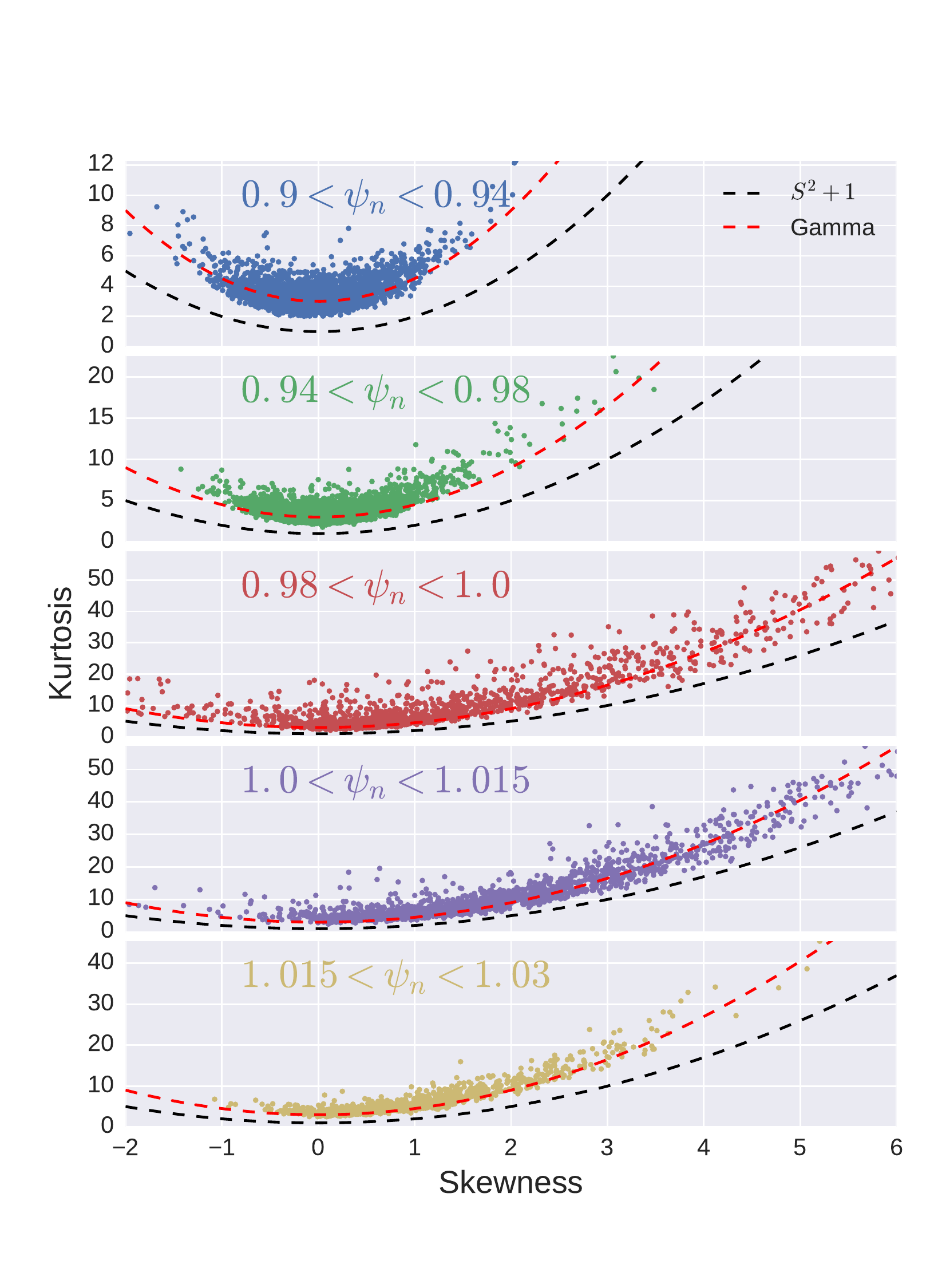}
\caption{Skewnesss vs kurtosis}
\label{fig:skewvskurt}
\end{figure}

Comparing the kurtosis as a function of the skewness gives an indication of the underlying distribution function\cite{Labit2007,Krommes2008}. In Figure \ref{fig:skewvskurt}, we see a general trend of gamma (or beta) distributed fluctuations ($K  = 1.5 S^2 + 3$) across the entire pedestal/SOL region. This suggests that these turbulent structures are transported nonlocally through the entire edge region.

\subsection{Conditional Spectrum}\label{sec:cond}
\begin{figure*}
\centering
\includegraphics[width=\textwidth]{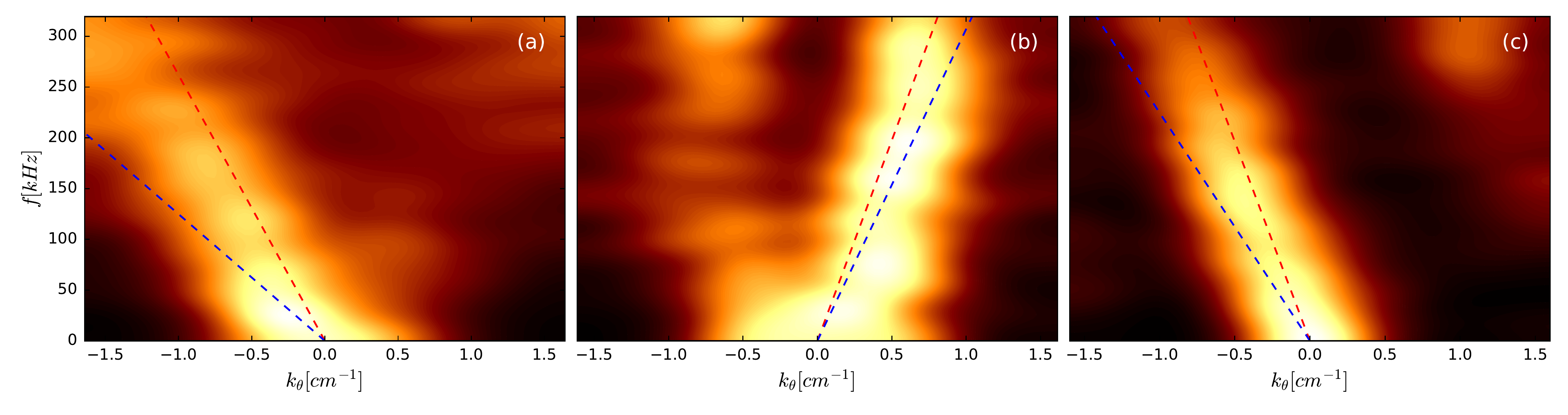}
\caption{Conditional spectrum at $\psi_N=0.91$, 0.98, and 1.02. Dashed red lines are a linear fit to the dominant mode. Blue dashed lines are the $E \times B$ velocity.}
\label{fig:kfspecs}
\end{figure*}

The conditional spectrum $S(k_\theta | f)$ has been utilized in numerous experiments \cite{Ritz1984,Endler1995,Cziegler2011} to identify unique features in fluctuating plasma quantities, such as the dominant turbulence mode (e.g. ITG, TEM, etc.). Here we use the LFS wedge region defined in the previous section, and calculate the conditional spectrum on each flux surface as:
\begin{equation}
S(k_\theta | f) = \frac{S(k_\theta,f)}{\sum_{k_\theta} S(k_\theta,f)}
\end{equation}

where $S(k_\theta,f)$ is the 2D Fourier transform over the poloidal coordinate $L_\theta$ and time. Here we preprocess the fluctuating electron density $\delta n_e$ with a Hanning window to avoid aliasing effects. The result is summarized in Fig. \ref{fig:kfspecs}.

As expected, the spectrum shows that the dominant turbulent mode across the pedestal and into the scrape-off layer has a lab-frame velocity close to the background $E \times B$ flow. However, there is a transition region in the location of the steep radial electric field well where dual modes exist, with lab frame phase velocities in opposite directions. This is the same observed by the BES diagnostic at DIII-D (although for L-H discharges)\cite{Yan2011}. Physically, this dual structure corresponds to a transition region of sheared turbulence, with some turbulent eddies moving upwards and others downwards.

\begin{figure}
\centering
\includegraphics[width=0.45\textwidth]{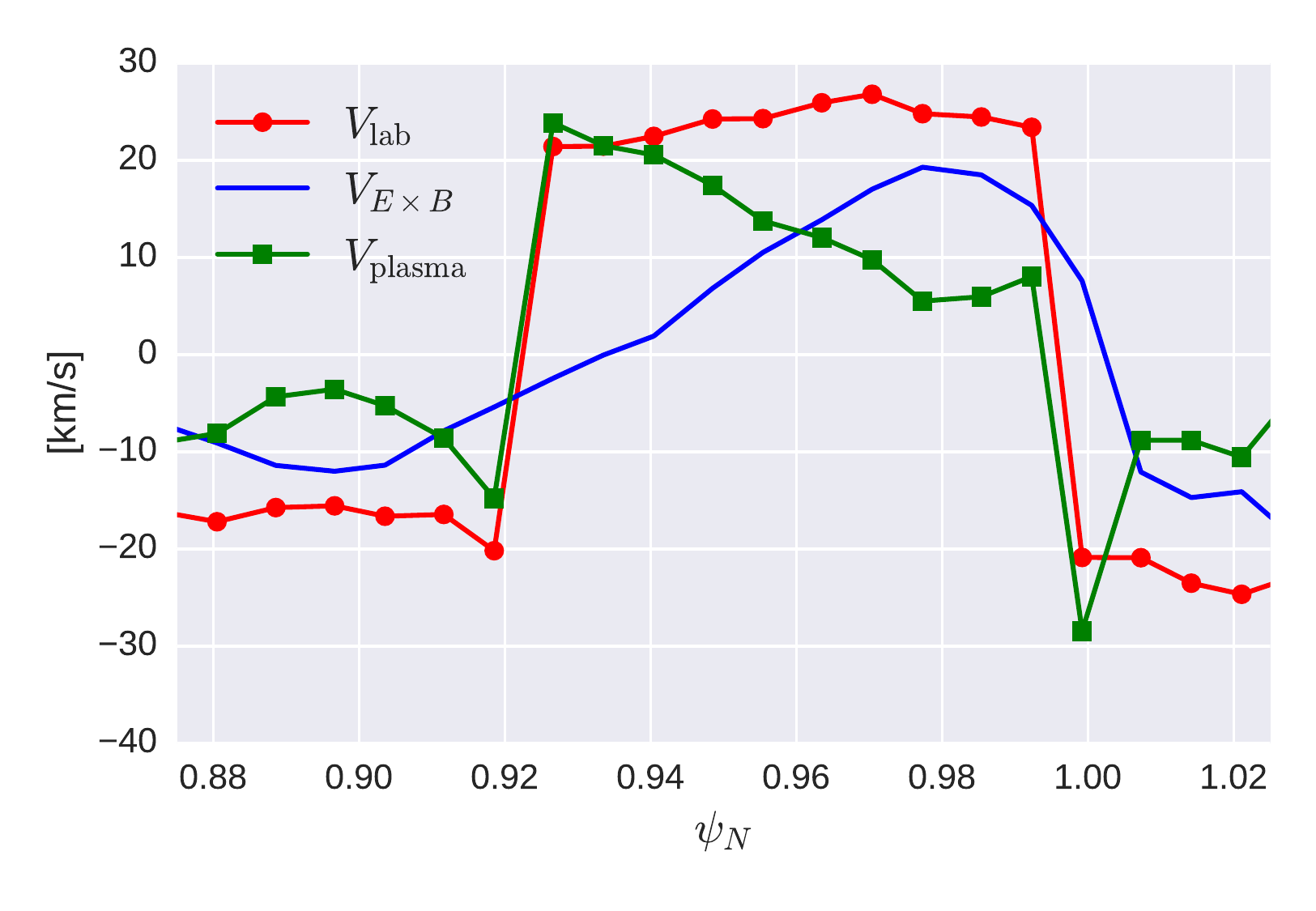}
\caption{Lab (or measured) phase velocity (red circles) derived from the conditional spectrum, $E \times B$ velocity (blue line), and the plasma phase velocity (green squares), formed by the difference between the two.}
\label{fig:velocity}
\end{figure}

A line was fit to the dominant mode (red dashed line in Figure \ref{fig:kfspecs}), and its average phase velocity in the lab frame calculated using the slope of the line ($V_{lab} = 2\pi f_{dominant} / k_{\theta,dominant}$). The local $E \times B$ velocity is depicted as a dashed blue line. These velocities, along with the mode phase velocity in the plasma frame, created from the difference $V_{plasma}=V_{lab} - V_{E \times B}$, are shown at several flux surfaces in Figure \ref{fig:velocity}. The positive direction is the electron diamagnetic direction (EDD), and the negative direction is the ion diamagnetic drift direction (IDD). Since this is an electrostatic simulation, with ion scales, the dominant turbulent modes we can identify according to the mode plasma frame phase velocity direction (green squares in Fig. \ref{fig:velocity}) is the ion temperature gradient instability (ITG) at $\psi_N<0.92$ and trapped electron mode (TEM) at $0.92<\psi_N<0.98$. However, similar to the sharp transition to the electron diamagnetic direction at $\psi_N=0.92$, there is another sharp transition to the ion diamagnetic drift direction in the lab-frame velocity crossing the separatrix, which appears to be responsible for the blobby turbulence dynamics, and may involve an instability different from the ITG. Since we have not turned on Coulomb collisions, this cannot be from the low-$\beta$ electrostatic-branch resistive ballooning mode activities, as claimed by fluid simulations\cite{DIppolito2011}, and a new interpretation is needed.  Density gradient driven drift modes associated with high safety factor $q$ are more likely the candidates. A more detailed investigation will be performed before we make a new report on this issue. We note that these sharp transitions, and the double-propagating feature, are similar to results from an Ohmic L-mode in Alcator C-Mod\cite{Cziegler2011}.

Another special observation can be noted here.  When we compare the mode phase velocity $V_{lab}$ in the laboratory frame (red circles) with the $E \times B$ velocity,  $V_{E \times B}$ (blue line), $V_{plasma}$ is in the in the electron direction for the ITG mode (Fig. \ref{fig:kfspecs}a) and the ion diamagnetic direction for the trapped electron mode (Fig. \ref{fig:kfspecs}b). This observation is in contradiction with the conventional understanding that the mode velocity direction should be identified relative to the $E \times B$ speed that was originated from the MHD type of ordering where $E \times B$ speed is much larger than the diamagnetic speed. Around the steep pedestal, the $E \times B$ speed roughly cancels the ion diamagnetic drift to the lowest order and the remaining fluid speed is smaller that the $E \times B$ speed. This is another subject to be studied in the near future.

\section{Blobs}\label{sec:blobs}
There is abundant experimental evidence that turbulence in the edge forms coherent ``blobs" of increased density that are sheared off from turbulence eddies, and propagate into the SOL\cite{DIppolito2011}. The importance of this convective transport has been estimated to be as much as 50\% of the total transport crossing the separatrix into the SOL\cite{Boedo2003}. These blobs are generally believed to propagate by $E \times B$ motion, due to a dipolar potential which forms in the blob, carrying particles and heat out of the plasma. Blob characteristics and mechanisms for blobs to transport into the SOL have been extensively studied, but much work is needed to understand the blob generation mechanism\cite{Krasheninnikov2008a}.

Here we investigate the characteristics of blobs created in this XGC1 simulation. Since XGC1 is a gyrokinetic code, these blobs are not seeded in the simulation but are rather generated by physical processes in the confined plasma region.

\subsection{Blob detection and tracking}
A blob detection and tracking method was applied to this XGC1 simulation, using a contouring algorithm to identify coherent blob structures, based off algorithms used in the NSTX experiment\cite{Davis2014,Zweben2015}. This method was applied to each poloidal plane, and the resulting velocities of blobs determined by the distance moved between time frames, and decomposed into radial and poloidal directions. Ellipses were also fit to the contours, to extract the radial and poloidal size of the blobs, and their area. The criterion for the value of $n_e / \langle n_e \rangle_\phi$ which indicated a blob is selected based on the standard deviation of density fluctuations in the region of strongest blob activity, in order to isolate intermittent structures. For this simulation, the criteria of $n_e/\langle n_e \rangle_\phi>1.2$ was used.

For the LFS region defined previously (see Section \ref{sec:turb}), the detected average radial blob size was 1.06 cm, and the average poloidal blob size was 4.6 cm (using the contour directly at $n_e/\langle n_e \rangle_\phi=1.2$), fairly close to the correlation lengths from Section \ref{sec:turb}.

\begin{figure}
\centering
\includegraphics[width=0.45\textwidth]{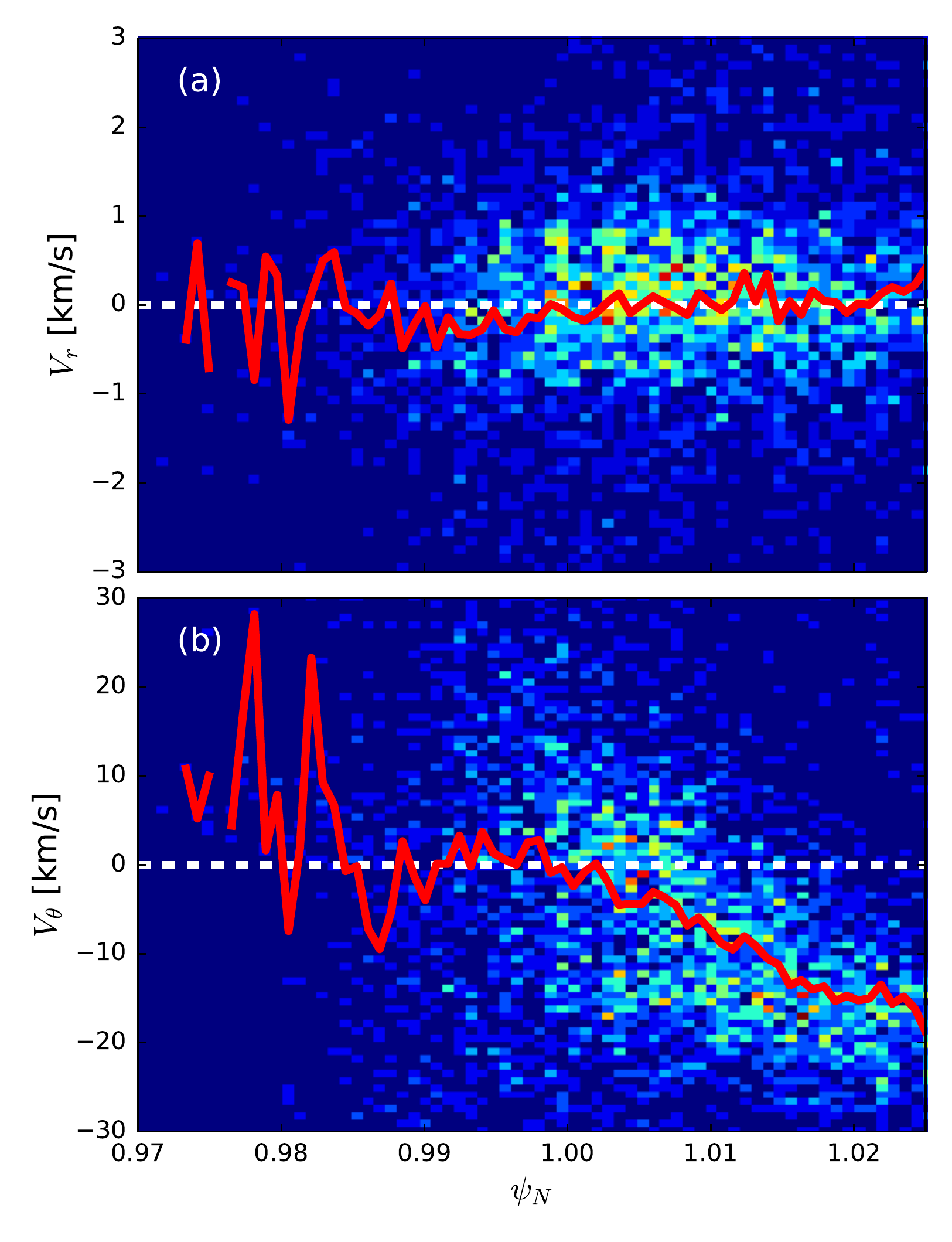}
\caption{2D histogram of blob velocities (color plots) with the mean overplotted in red (a) radial velocity and (b) poloidal velocity}
\label{fig:blob_vels}
\end{figure}

The blob velocities derived from the tracking algorithm are shown in a 2D histogram in Figure \ref{fig:blob_vels}, along with the mean velocity overplotted. Blobs have a dominant positive poloidal velocity (EDD) in the confined region ($\psi_N<1$), and switch to the negative direction (IDD) in the SOL. The radial velocity of the blobs show a large scatter around zero, with the standard deviation being about $\pm 1$ km/s. The mean slightly increases to a few tenths of km/s further into the SOL. A similar scatter in the radial velocity of blobs in H-mode were observed on NSTX\cite{Zweben2016}.

\begin{figure}
\centering
\includegraphics[width=0.45\textwidth]{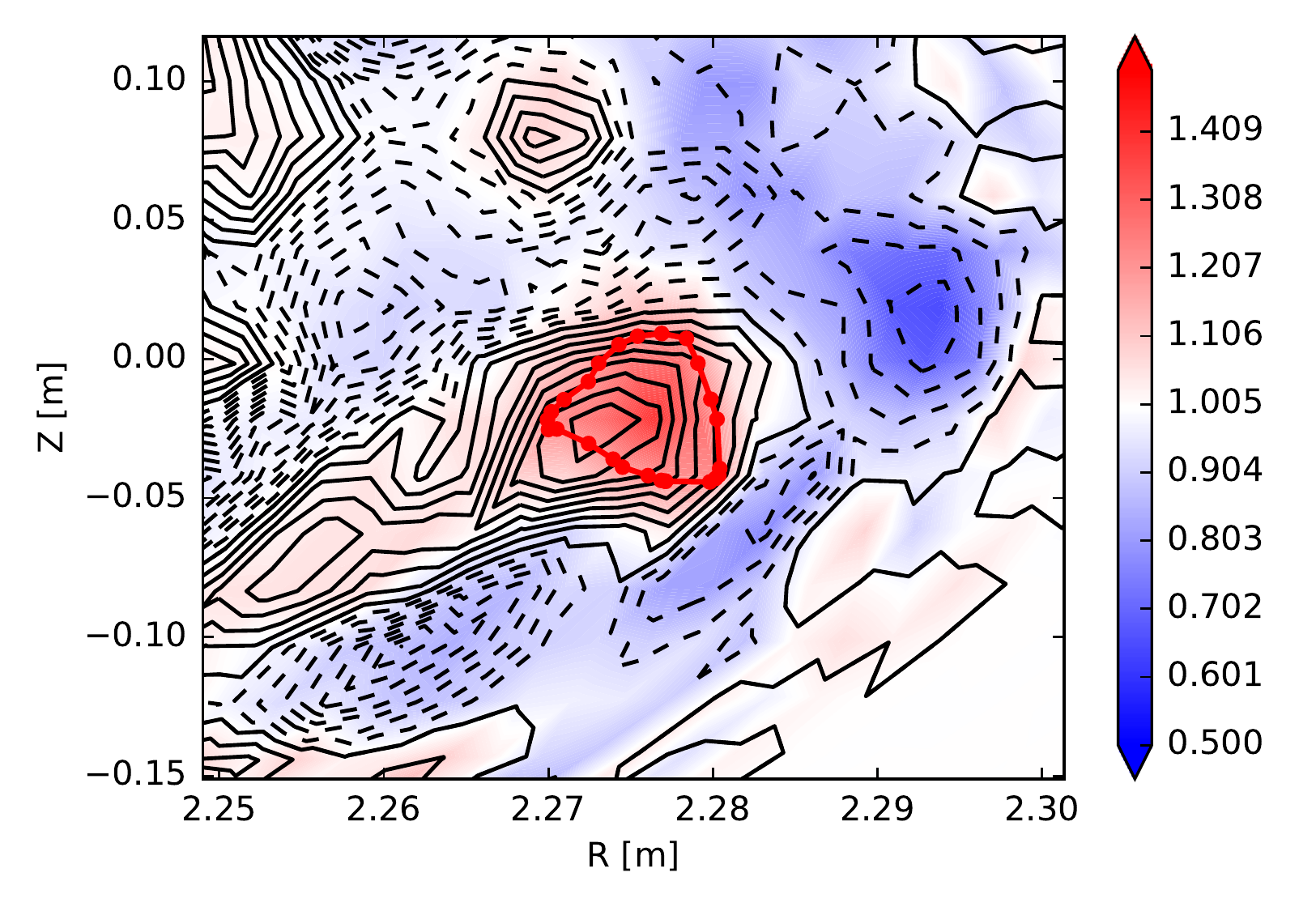}
\caption{Normalized density ($n_e / \langle n_e \rangle_\phi$) at a single poloidal plane, in color (scale indicated in the colorbar), with fluctuating potential contour lines overlaid to show positive potential (solid black) and negative potential (dashed black). A detected blob is outlined in the thick red contour line. }
\label{fig:blob_potential}
\end{figure}

\begin{figure}
\centering
\includegraphics[width=0.45\textwidth]{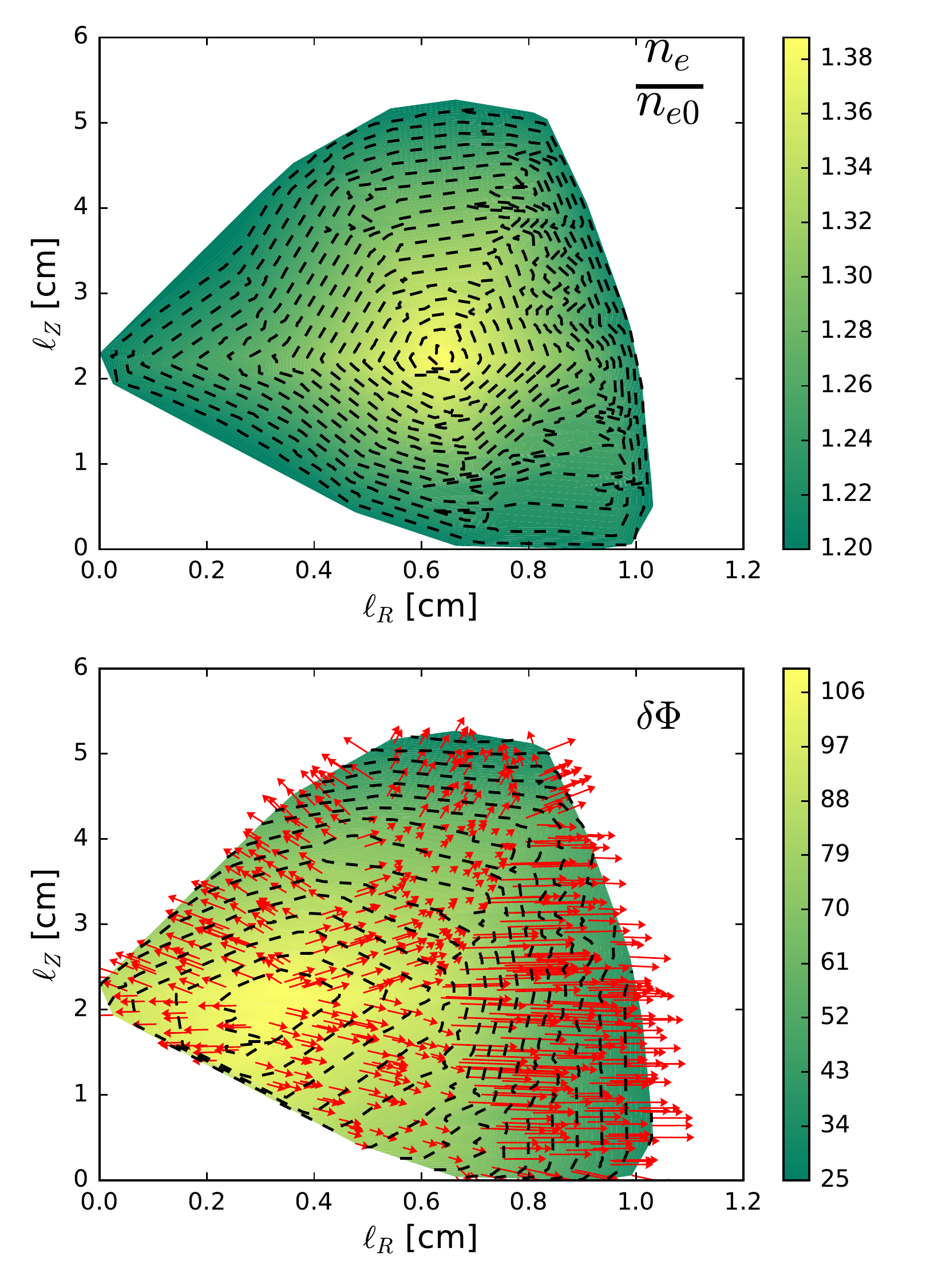}
\caption{Density and potential inside the blob outlined in the thick red line in Figure \ref{fig:blob_potential}, $R=2.275$m and $Z=-0.025$m. Red arrows in the potential plot show electric field. $\ell_R$ is the major radius distance, and $\ell_Z$ the vertical distance of the blob.}
\label{fig:blob_ne_vs_potential}
\end{figure}

The common picture of blob motion involve a charge polarizing force, in tokamaks due to magnetic drifts, which gives rise to a poloidal dipolar potential structure within the blob, causing the blob to move radially by $E \times B$ motion\cite{Krasheninnikov2001,DIppolito2011}. Certain mechanisms such as sheared flow can rotate this dipole to give rise to partially poloidal motion\cite{Myra2013}, and even give rise to monopole components causing blob rotation. As seen from Figure \ref{fig:blob_vels}, the average blob motion has a strong poloidal component, begging the question of what the potential structure looks like inside these blob. Figure \ref{fig:blob_potential} shows a filled contour plot at the LFS midplane of the normalized density, $n_e / \langle n_e \rangle_\phi$, with black lined contours overlayed to show the positive (solid black) and negative (dashed black) fluctuating potential, $\delta \Phi = \Phi - \langle \Phi \rangle_\phi$. A dark solid red contour outlines a detected blob. As seen, the potential structure is not dipolar but has a shifted potential pole peak, with the potential inside the blob being completely positive. This can be seen even clearer in Figure \ref{fig:blob_ne_vs_potential}, where side by side plots of the blob's normalized density and potential are shown. The blob density is clearly monopolar, while the potential is also monopolar but with a shifted peak relative to the density peak. Figure \ref{fig:blob_ne_vs_potential} also shows the electric field vector at several points inside the blob (calculated at random location using gradients on 2d linear interpolations of potential), indicating a dominantly radial electric field across the flux-surfaces, which would lead to dominantly poloidal $\mathbf{E} \times \mathbf{B}$ motion, together with blob rotation. This analysis shows that the simple up-down charge-separation argument is not always present with blob creation, and that the average predominant blob motion is not always purely across the flux surfaces.

\section{Discussion}\label{sec:discussion}
The electrostatic turbulence and blob characteristics from this collisionless XGC1 simulation discussed in this paper give insight into the dynamical system of the plasma edge. However, here we have only scratched the surface of the insight to be provided by these large, multiphysics XGC1 simulations in realistic geometry. Collisional dissipation could also alter the physics discussed here.

Of particular interest is understanding at a fundamental level the blob generation mechanism\cite{Krasheninnikov2008,Krasheninnikov2008a}, which due to the difficult nature of the task has received less investigation by the community than blob transport. The topic of blob generation is of particular importance for predicting the divertor heat flux width on ITER, as recent XGC1 simulations suggest the relative importance of blobby to neoclassical transport leads to a larger (i.e. safer) divertor heat load width on ITER compared to extrapolations using current-day experiments\cite{Chang2017}.

Additionally, the realistic geometry will allow further investigations in particular of the effect of the X-point on blob dynamics. Already it is observed in this simulation that the blobs elongate radially near the X-point as expected\cite{Krasheninnikov2008a}, due to the flux expansion there. But they do not travel much further than the X-point, dying out before reaching the LFS divertor plate. This is important as many theories assume the blobs are connected to the divertor plate, which greatly affects the parallel current flowing through the blob, an important piece of blob sustainment. Future work can isolate and determine the role the X-point has on blob dynamics, including comparing to analytic models of X-point effects\cite{Myra2013}.

The realistic geometry also allows studying how 3D effects affect blobs. Prototype algorithms have been created and applied for tracking XGC1 blobs in 3D and the PIC particles that pass through them\cite{Pugmire2016}. Such tracking in 3D will give information on blob parallel wavelength, and parallel variation of plasma parameters within the blob.

Finally, the role of ion temperature on the blobs should be investigated more closely with these XGC1 simulations. Analytically it is known that ion temperature has a host of effects on blob transport, including adding poloidal motion\cite{Jovanovic2008}. A recent XGC1 simulation showed reduced blob generation when using an input $T_i$ profile that decreased mainly in the SOL. This underlines the importance of quality main ion temperature measurements in both the pedestal and SOL, which until recently relied on a proxy of impurity ion temperature. It is now being shown that $T_z \ne T_i$ in the pedestal region\cite{Haskey2016}, with the SOL an open question (SOL ion temperature measurements, main ion or impurity, are still not reliably made). How $T_i$ affects blob generation and transport in these XGC1 simulations is an open question.

\section{Conclusion}\label{sec:conclusion}
The electrostatic turbulence and collisionless blob characteristics in the edge (pedestal + SOL) of an XGC1 simulation of a DIII-D-like H-mode plasma have been detailed. Normalized density fluctuation levels increase around the maximum in the electron pressure gradient, and stay high into the SOL. Intermittent activity is visible in time traces, and in the skewness and kurtosis which follow a similar trend as the normalized fluctuation level. It was shown that throughout the pedestal and SOL the density fluctuations follow a gamma distribution skewness/kurtosis relationship. Conditional spectra showed the $k_\theta -f$ characteristics, with double lobed features in regions of sheared flow, and wave velocities consistent with the drift-wave ITG instability near the top of the density pedestal, and TEM in the mid to bottom of the pedestal. Electron pressure driven drift wave type turbulence then give rise to blobs from $\psi_N > 0.98$. A blob tracking algorithm was used to isolate coherent spatial blob structures, showing blob radial and poloidal size comparable to the correlation lengths. Blob radial velocities showed a large scatter, contained mainly within $\pm 1$ km/s, with the average near zero, and increasing somewhat into the positive direction in far-SOL. The poloidal blob velocities average decreases further into the SOL, reaching $-20$ km/s in the far-SOL. The potential distribution within blobs is generally not dipolar, but rather a shifted monopole, with a significant $E_r$ leading to poloidal $E \times B$ motion of the blob.

Understanding from a fundamental level the turbulence and transport processes in the edge region will benefit the fusion community as it looks towards ITER and beyond. The multi-physics, whole-device goal of XGC1 will aid in this process, and future work in identifying, isolating, and extracting physics insight from XGC1 simulations will be paramount to this end.

\section{Acknowledgements}
The authors would like to thank Dr. Stewart Zweben for his insightful comments on the manuscript. This work supported by the U.S. Department of Energy SciDAC program under DE-AC02-09CH11466, DE-AC05-00OR22725, and DE-FC02-04ER54698.  Awards of computer time was provided by OLCF through the Innovative and Novel Computational Impact on Theory and Experiment (INCITE) program. This research also used resources of the National Energy Research Scientific Computing Center, a DOE Office of Science User Facility supported by the Office of Science of the U.S. Department of Energy under Contract No. DE-AC02-05CH11231.

\section*{References}

\bibliography{turbulence_paper}

\end{document}